# An ionization sensor scheme for ultra-low voltage operation using one-dimensional nanostructures

Zhongyu Hou[1]*, Maobo Fang

[1]*National Key Laboratory of Micro/Nano Fabrication Technology, Key Laboratory for Thin Film and Microfabrication of Ministry of Education, Research Institute of Micro/Nano Science and Technology, Shanghai Jiao Tong University, Shanghai 200030, People's Republic of China.*

**ABSTRACT**

The one-dimensional nanostructures have been used as the electrode to decrease the operation voltage where strong electric field is needed to function the device in the gaseous electronics. In this letter, a novel electrode scheme is proposed to generate high field intensity at extremely low applied voltages. Through the theoretical calculation, we shall demonstrate some specific cases of the electrode systems where the fields can reach $10^6$, $10^9$, and $10^{11}$ V/m orders of magnitudes at the applied voltage as low as 10mV. The imaginary cases could be realized by various fabrication technologies.

---

1  Corresponding author: zhyhou@sjtu.edu.cn

*Introduction*: The exercise of using one-dimensional nanostructures (ONSs) as the electrode surface modification has been among the major focuses of the cutting-edge studies in the gaseous electronics. The dedicated literatures may be classified into four different contexts: the ionization gas sensors (IGSs) [1-5], field ionization sources (FISs) [6-8], micro-plasma devices [9,10], and fundamental phenomena and mechanisms in gaseous electronics [11-13].

The author would like to share a novel scheme of utilization of the ONSs in an electrode system to generate the intensive electrostatic field at ultra-low applied voltages, and also the time-variant electric field. The essential character of this methodology is to position the ONSs near the charge sources but without direct electric contact with any of them. An on-chip-device oriented example is to configure the ONSs between two metallic pads of different electric potential. One can imagine many other interesting scenarios to take the advantages of this methodology; however, in the following, the author would focus on the specific case above-mentioned.

*Electrostatic field calculation:* As shown in Fig. 1, an array of one-dimensional electrodes is positioned between two electrode pads. When an electric potential difference is applied, electric field in the region between those two pads will polarize the ONSs. If the ONSs are metallic, electrons will concentrate at the surface of the ONSs and form an equivalent dipole moment so that the field inside the ONSs is balanced out. To calculate the field distribution, we first determine the induced surface charge density of the ONSs, and numerically calculate the field distribution through finite element method. The method is also valid if the ONSs are semiconductors; however, the determination of the surface charge density of the ONSs will be different.

Supposed that the specific structure of the electrode in the ONSs is the rod in nanoscale. The equivalent surface charging density, $\sigma_p$, could be given by:

$$\sigma_p = \frac{\varepsilon_0 h_e M R}{(A_e)^2} \left[ \frac{2A_e}{a} + \frac{M(N-1)Rh_e}{s} + \frac{(M-1)sh_e}{d-2a} \right] \cdot V, \tag{1}$$

where $M$ and $N$ are the numbers of the nanorods in the direction normal to and parallel with the electrode plate surface, facing to the gap, respectively; $R$ is the averaged equivalent radius of the nanorods, $A_e$ is the area of the electrode surface, $h_e$ is the height of the nanorods, $d$ is the gap spacing, $a$ is the average distance between the nanorods area and the electrode surface, and $V$ is the applied voltage. The model has ignored the conduction of the charges among different inter-connected nanorods, which means that the equation is best fitted for the metallic nanostructures with perfect electric isolation.

Based on equation 1, the surface charge density can be very high even when $V$ is very low, so that the field intensity can also be controlled in the range of $10^{11}$V/m order of magnitude. In this scenario, the methodology could be applied as a perfect field ionizer with very low applied voltage. For example, if $R$=10nm, $a$=300nm, $N$=300, $M$=1500, $A_e$=0.8μm², $s$=600nm, $h_e$=1μm, $d$=2$a$+$M$($s$+$R$), when $V$=10mV, $\sigma_p$ can be about 6.2×10⁻¹C/m²; this corresponds to the field intensity at the edge point area in the order of 2.62×10¹¹V/m, according to a finite element method solution of the Laplacian equation. For another example, if $R$=200nm, $a$=200nm, $N$=10, $M$=10, $A_e$=0.297μm², $s$=2μm, $h_e$=20μm, when $V$=10mV, $\sigma_p$ can be about 8.0×10⁻³C/m²; this corresponds to the field intensity at the edge point area in the order of $2.67\times10^9$V/m. For the third example, if $R$=10nm, $a$=300nm, $N$=100, $M$=100, $A_e$=0.8μm$^2$, $s$=600nm, $h_e$=1μm, when $V$=10mV, $\sigma_p$ can be about $9.92\times10^{-4}$C/m$^2$; this corresponds to the field intensity at the edge point area in the order of $2.06\times10^8$V/m and the channel regions between every two electrode pair is in the order of $10^6$V/m. In this scenario, the gas breakdown is expected in the channel regions.

*Remarks*: The realization of some of the geometrical configuration is only possible when the fabrication methodology has very high precision in nanometer scale. Perhaps, the controlled fabrication methods, e.g., the focused ion beam, is among the best choices. For the device fabricated through chemical reaction methods, the inter-connection among different nanorods cannot be ignored, the conduction between the electrodes will cause the leakage of the polarization-induced charge accumulation. In this case, the utilization of equation 1 will overestimate the realistic field intensity. Besides, if the ONSs are semiconductors or insulators, equation 1 is not applicable also. Future works will be shared in this forum, recently.

Another specific property of this scheme: one can realize different charge separation processes, say, field ionization, field emission, and electron avalanche, through proper control of the geometrical feature of the polarization nanostructures. For example, if the one-dimensional nanostructures are thin enough so that the field enhancement effect is highly localized, the establish of a higher electric field is still possible because the discharge current can be controlled in low level.

The authors welcome every scientific criticism.

This work and the future works refer to a Chinese patent which is suspended for authorization and will be published in Chinese soon.

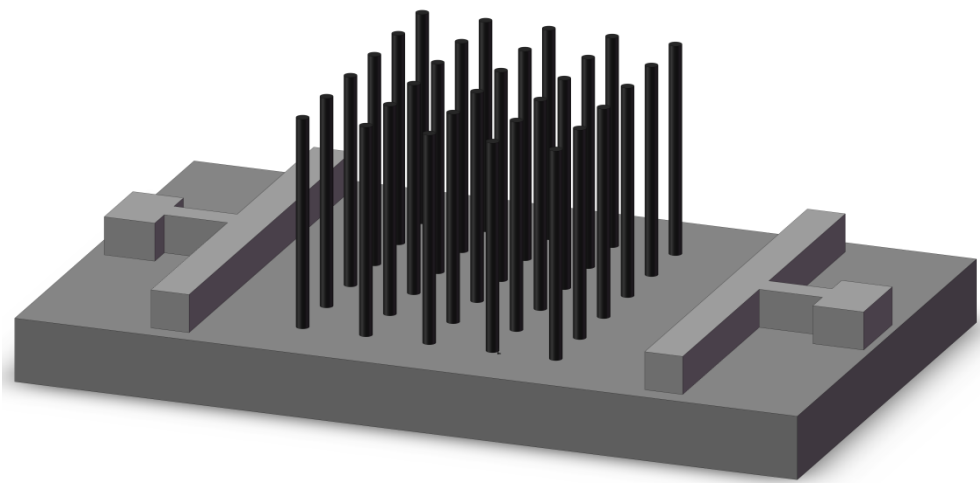

Fig. 1 The schematic of the scheme.